\documentclass[astron]{eas}
\usepackage{graphicx}
\pdfoutput=1
\def\lesssim{\mathrel{\hbox{\rlap{\hbox{\lower4pt\hbox{$\sim$}}}\hbox{$<$}}}}
\def\gtrsim{\mathrel{\hbox{\rlap{\hbox{\lower4pt\hbox{$\sim$}}}\hbox{$>$}}}}
\begin{document}

%%-----------------------------
%%      the top matter
%%-----------------------------
\title{Supershells in the Multi-Phase Milky Way: Insights from H{\sc i} Synthesis Imaging and CO Surveys} 
\runningtitle{Supershells in the Multi-Phase Milky Way}
\author{J. R. Dawson}\address{Department of Physics and Astrophysics, Nagoya University, Chikusa-ku, Nagoya, Japan; \email{joanne@a.phys.nagoya-u.ac.jp \&\ fukui@a.phys.nagoya-u.ac.jp}}
\author{N. M. McClure-Griffiths}\address{Australia Telescope National Facility, CSIRO, Marsfield, NSW 2122, Australia; \email{naomi.mcclure-griffiths@csiro.au}}
\author{Y. Fukui}\sameaddress{1}
\begin{abstract}
We present new parsec-scale resolution data from a multi-phase study of the ISM in the walls of Galactic supershells. H{\sc i} synthesis images and CO survey data reveal a wealth of substructure, including dense-tipped fingers and extended molecular clouds embedded in shell walls. We briefly consider formation scenarios for these features, and suggest that both the interaction of an expanding shell with pre-existing dense clouds, as well as in-situ formation of CNM and molecular gas, are likely to be relevant. Future work will also examine the role of instabilities in structure formation and breakup, and will investigate the presence of high-altitude gas associated with supershells and chimneys.  
\end{abstract}
\maketitle
%%-----------------------------
%%      your text
%%-----------------------------
\section{Introduction}

Supershells and chimneys formed by stellar feedback are a ubiquitous presence in disk galaxies. They strongly impact the structure and evolution of the disk ISM, and play a crucial role in the exchange of energy and mass between the disk and the halo. 

Here we choose to focus on the role played by supershells in the formation, destruction and transportation of the cold neutral medium (CNM) and molecular clouds. Supershells form part of an inhomogeneous, richly-structured and multi-phase ISM, and an understanding of the processes that dominate their evolution in this complex medium is needed if we are to construct a comprehensive theory of the disk-halo interaction. The accumulation of the ISM in expanding shells is an effective way of generating cool, dense high-column density conditions over large spatial scales; conditions that naturally favour the formation of CNM and molecular gas. 
Yet conversely, the interaction of a supershell shock front with pre-existing dense clouds may also be a disruptive influence, acting to tear them apart. 
The interplay between these processes and the development of instabilities in the expanding shell is also likely to be important, but remains poorly understood. 

The present work takes some first steps towards an observational investigation of these issues,
via the study of the neutral ISM in Galactic supershells. We utilise H{\sc i}  synthesis images from the Australia Telescope Compact Array (ATCA), combined with Parkes 64m single dish data. With these data we achieve unprecedented parsec-scale resolutions, revealing detailed substructure in atomic shell walls. In addition, unique to this study is the use of matched-resolution $^{12}$CO(J=1-0) data, taken with the NANTEN 4m telescope. This powerful multi-phase approach enables us to pinpoint the locations of the cold, dense molecular phase of the ISM within our shell systems.\par 

This paper deals primarily with two objects for which the best quality data exist - GSH 287+04-17 (Dawson {\em et al.\/} \cite{dawson08}) and GSH 277+00+36 (McClure-Griffiths {\em et al.\/} \cite{mccg03}). However, preliminary examination of lower resolution data suggests that the features discussed below may be common to many Galactic Supershells. 

\section{Cold, Dense-Tipped Fingers}
\label{fingers}

The data reveal striking finger-like protrusions in the walls of our sample supershells. These point radially inwards towards the shell interior, with the main walls pulled back, arch-like, around them. Examples of these features can be seen in Figure 1, which shows a subsection of the wall of GSH 287+04-17.\par 

Atomic gas temperatures in these fingers are remarkably cold. H{\sc i}  linewidths in their tips fall between $2.5 < \Delta v < 5.0$ km s$^{-1}$, which corresponds to kinetic temperatures of $150 < T_K < 550$ K under the assumption of pure thermal broadening. In reality, turbulence is likely to contribute significantly to the observed linewidths and these figures are strict upper limits. 
The $^{12}$CO(J=1-0) data reveal molecular clouds with masses of $10^{2-3} M_{\odot}$ at the tips of some -- but not all -- of the atomic protrusions. This molecular gas is in all cases co-moving (in $v_{lsr}$) with the spatially coincident H{\sc i}, and the integrated CO luminosity implies column densities of $N(\mathrm{H}_2)\approx5$-$50 \times 10^{20}$ cm$^{-2}$. Yet, other than the presence or absence of CO, we find no significant difference between the properties of the ISM in the molecular and non-molecular protrusions. No obvious trends are present in H{\sc i} morphology, linewidths, intensities or implied column densities. It is of course possible that H{\sc i} and CO alone fail to effectively trace the entire neutral ISM column density, and we do not rule out the possibility that H$_2$ gas with low CO luminosity may be present even in the ``non molecular'' fingers.

The morphology of these fingers is suggestive of the interaction of the expanding shell with pre-existing dense clouds.
We may perform a simple check on this scenario by comparing theoretical cloud destruction timescales with the molecular cloud survival times implied by the data. Assuming that the passage of the shock has not accelerated our clouds significantly, and ignoring projection effects, we use measured shell expansion velocities and cloud-to-main-wall distances to obtain a lower limit of several $10^6$ yr for the cloud survival time. We envision a situation in which post-interaction clouds are left sitting in a bath of hot, tenuous interior gas, and consider both classical evaporation and shock destruction timescales. For clouds of $n_{c} \sim 100$ cm$^{-3}$, $R_{c} \sim 2$ pc embedded in a hot interior medium of $n_i \sim 0.01$ cm$^{-3}$, $T_i \sim 10^6$ K, the evaporation timescale is given by $\tau_{evap} \sim 3.3\times10^{20}~n_cT_i^{-5/2}R_c^2 \sim 10^8$ yr (Cowie \& McKee \cite{cowie77}). The shock destruction timescale may be expressed in terms of Òcloud crushing timeÓ, $\tau_{cc}$, as $\tau_{dest} \sim 10~\tau_{cc} = 10~(R_c/v_{exp}).(n_c/n_0)^{0.5}$ (Klein {\em et al.\/} \cite{klein94}; Silich {\em et al.\/} \cite{silich96}; Nakamura {\em et al.\/} \cite{nakamura06}). For a cloud impacted by a shell with $n_0\sim5$ cm$^{-3}$, $\tau_{dest} \sim 10^7$ yr. This comparison is necessarily quite crude, and is dependent on the choice of initial parameters, as well as on initial model setup. Nevertheless, for realistic parameter sets we obtain figures that are comfortably consistent with the survival of the observed clouds, while not being so long as to suggest destruction is ignorable over the shell lifetime. 

\begin{figure}
\begin{center}
\includegraphics[scale=0.85]{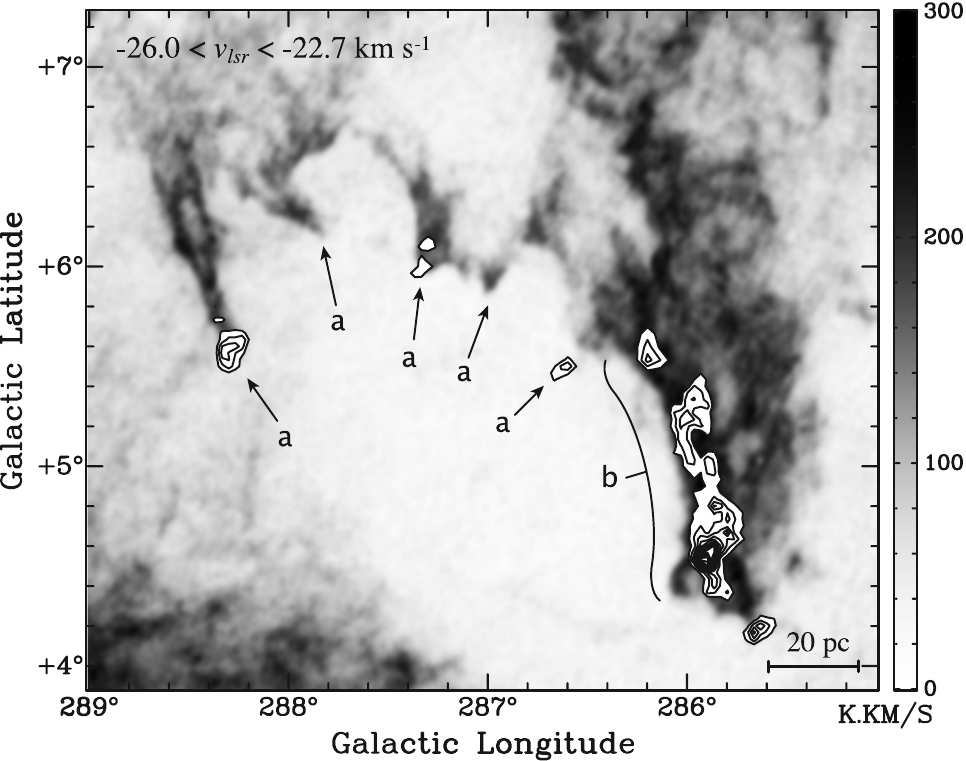}
\caption{Substructure in the walls of GSH 287+04-16. The greyscale image shows combined ATCA and Parkes H{\sc i} data. Filled contours show NANTEN $^{12}$CO(J=1-0) data. Features labelled ``a'' are discussed in section \ref{fingers}. The feature labelled ``b'' is discussed in section \ref{embedded}.
}
\end{center}
\end{figure}

\section{Embedded Molecular Clouds}
\label{embedded}

As well as dense-tipped fingers, we also observe extended complexes of molecular clouds coincident with the main shell walls. These cloud complexes have masses of $M(\mathrm{H}_2)\sim10^4 M_{\odot}$, and are associated with comparable masses of H{\sc i}. An example may be seen in Figure 1. 

In contrast to the H{\sc i}/CO fingers, the morphology of these features is strongly suggestive of dense cloud formation from the swept-up medium. Making some simple assumptions about geometry, we may estimate the average number density in the ambient medium that would be required produce the observed clouds. We find $\langle n \rangle=1-10$ cm$^{-3}$ for the two shells examined in this study, which is plausible in a realistic inhomogeneous ISM. Recent numerical work suggests that the formation times for CO-rich molecular gas from weakly supersonic atomic flows is $\sim1\times10^7$ yr for initial number densities of $n=3$ cm$^{-3}$ (Bergin {\em et al.\/} \cite{bergin04}, Heitsch \& Hartmann \cite{heitsch08}). This figure is consistent with the ages of $\sim10^7$ yr estimated for the two shells examined in this study. Therefore molecular cloud formation would appear to be plausible of over the shell lifetimes.   

\section{Summary and Future Directions}

We are using new, high-resolution multi-phase data to investigate the role played by supershells in the formation, destruction and transportation of the CNM and molecular clouds. The preliminary analysis described above suggests that encounters with small pre-existing dense clouds will disrupt them on timescales comparable to the shell lifetime, and that such encounters can therefore provide an explanation for the cold fingers observed protruding into supershell interiors. In contrast, extended molecular complexes apparently embedded in shell walls can be plausibly formed from swept up WNM over typical supershell lifetimes.

Future work will examine the role of instabilties,
motivated in part by the periodic, ``scalloped'' morphology observed some Galactic supershells. We also note that unusually high altitude molecular clouds are associated with some shells. Whether such clouds were transported from low latitudes or formed it situ from the swept up material, their existence suggests that supershells are capable of supplying cold dense gas to the highest regions of the Galactic disk. Investigation into both of these issues is expected to proceed in conjunction with numerical modelling.

%%-----------------------------
%%      your bibliography
%%-----------------------------

\end{document}